\documentclass[10pt]{IEEEtran}

\usepackage{graphicx}
\usepackage{algorithm,algpseudocode}
\usepackage{amsmath,amsfonts,amssymb,mathtools}
\usepackage{epsfig,amstext,xspace}
\usepackage{amsthm}
\usepackage[sans]{dsfont}
\usepackage{stmaryrd}
\usepackage{pifont}
\usepackage{wasysym}
\usepackage{array}
\usepackage{url}
\usepackage{color}
\usepackage{multirow}
\usepackage[skip=1pt]{subcaption}
\usepackage[belowskip=0pt,aboveskip=1pt, labelfont=bf]{caption}
\usepackage{enumitem}

\begin{document}


\title{Enabling Self-aware Smart Buildings by Augmented Reality}

\author{Muhammad Aftab, Sid Chi-Kin Chau, and Majid Khonji
\IEEEcompsocitemizethanks{\IEEEcompsocthanksitem M. Aftab is with Aalborg University. S. C.-K. Chau is with Australian National University. M. Khonji is with Masdar Institute, Khalifa University. 
E-mail: muhaftab@cs.aau.dk, chi-kin.chau@cl.cam.ac.uk, mkhonji@masdar.ac.ae
}
\IEEEcompsocitemizethanks{\IEEEcompsocthanksitem This paper appears in ACM International Conference on Future Energy Systems  (e-Energy), 2018
}}

\maketitle

\begin{abstract}
Conventional HVAC control systems  are  usually incognizant  of the physical structures and materials of buildings. These systems merely follow pre-set HVAC control logic based on abstract building thermal response models, which are rough approximations to true physical models, ignoring dynamic spatial variations in built environments. To enable more accurate and responsive HVAC control, this paper introduces the notion of {\em self-aware} smart buildings, such that buildings are able to explicitly construct physical models of themselves (e.g., incorporating building structures and materials, and thermal flow dynamics). The question is how  to enable self-aware buildings that automatically acquire dynamic knowledge of themselves. This paper presents a novel approach using {\em augmented reality}. The extensive user-environment interactions in augmented reality not only can provide intuitive user interfaces for building systems, but also can capture the physical structures and possibly materials of buildings accurately to enable real-time building simulation and control. This paper presents a building system prototype incorporating augmented reality, and discusses its applications.

\end{abstract}

%
%



\section{Introduction}

The rise of augmented reality provides an exciting venue for extensive human-machine interactions. In essence, augmented reality overlays computer-generated information with the physical world in a composite view of a user through motion tracking sensors and computer vision.  Entertainment, design and navigation are key applications of augmented reality \cite{van2010survey}. Nonetheless, the potential of augmented reality can also be harnessed by other innovative applications. In this paper, we explore a novel methodology utilizing augmented reality  as an empowering technological tool for more effective building automation, and particularly, HVAC control.

Traditional building management systems usually follow simple pre-set control logic programmed in the hardware controllers.  Often, a reductionistic approach is employed, wherein the thermal response behavior of buildings is approximated by first-principle models, and linear time-invariant dynamic processes with 1D parameters by ordinary differential equations (also known as lumped element resistance-capacitance (RC) models) \cite{aftab2013, mpc_rc_calib2, mpc_rc_calib1}. HVAC control strategies are then devised based on such abstract models, which can only  approximate simple building geometry in near-future time horizon. These first-principle models are often difficult to calibrate in a dynamic environment because the physical observable parameters from buildings (e.g., building structures, locations of doors and windows) are not explicitly incorporated in the models. Also, the error accumulates considerably when a longer time horizon of the control strategy is considered. While non-linear models are complicated and impractical, other alternatives based on true physical models of buildings are more viable. 

\subsection{Need for Self-aware Buildings}
To accurately and responsively control buildings, more precise representation of building thermal response behavior ought to be considered, particularly the detailed spatial-temporal dynamics of thermal flow.  The physical thermal response model of a building can be represented by partial differential equations, which often are employed in sophisticated building simulators \cite{simulators}. The conventional wisdom is that such building simulators need considerable processing power, and hence are only feasibly utilized in the design stage of buildings. Nonetheless, there have been considerable advances in embedded system technologies, which provide low-cost platforms with powerful processors and sizeable memory storage in a small footprint. There are opportunities to harness such embedded system technologies for real-time building automation systems. Particularly, accurate building thermal response simulations based on physical thermal response models can be executed efficiently in real-time on these embedded systems. Therefore, this paper introduces the notion of {\em self-aware} buildings, such that buildings can explicitly construct physical models of themselves (e.g., incorporating building structures, materials, and thermal flow dynamics) to enable accurate real-time simulation-guided control in building automation systems.

\subsection{Augmented Reality as an Enabling Tool}
Despite the promising benefits of self-aware buildings, the question is how  to enable self-aware buildings that can automatically acquire dynamic knowledge of themselves. This paper presents a novel approach using {\em augmented reality}. The extensive user-environment interactions in augmented reality not only can provide intuitive user interfaces for building management systems, but also allow the systems to capture the physical structures and possibly materials of buildings accurately. Augmented reality is also useful for eliciting interactive user feedback for personalization of building management systems. This paper presents a prototype of building management system incorporating augmented reality (based on Google's Project Tango platform) and discusses its applications.
{\color{black} Our main contribution is to demonstrate that augmented reality can help optimize building control by empowering building management systems to automatically recognize the building structures, and then infer the building material properties and HVAC system parameters.}
The advantages of incorporating augmented reality also include providing appealing motivation to users for upgrading their building management systems that can integrate with home infotainment systems.

Researchers are beginning to explore the use of augmented reality for building navigation, management, and control. The authors in \cite{diakite2016first} present opportunities and challenges involved in the use of 3D information of the indoor building environment for the purpose of indoor navigation. A method for 3D reconstruction of indoor and outdoor environments is proposed in \cite{schops20153d}. They perform several filtering steps on monochrome fisheye images to reconstruct scenes directly on a mobile device. A history of augmented reality technology, its challenges, and applications can be found in \cite{ar_review}.

\section{Self-aware Buildings}

Before defining the notion of self-aware buildings, we discuss several problems of traditional approaches \cite{mpc_rc_calib1, mpc_rc_calib2} that are based on abstract thermal response models for HVAC control. First, the true physical state of a building is approximated by a canonical state, which normally represents the average state in the space, without considering the spatial variations of thermal response behavior in different regions of the building. There is a considerable loss of state information in the abstract models, which are unable to incorporate the physical observable parameters from buildings (e.g., building structures, locations of doors and windows). Second, the measurements from the real world do not always match consistently with the outcomes of abstract models.  The lack of physical  structural information impedes the calibration of such models. Third, the dynamic process in an abstract model does not correspond to the true behavior of the physical world. As a result, the discrepancies of abstract models can grow considerably over time.

\begin{figure}[htb]
\centering
 \includegraphics[scale=1]{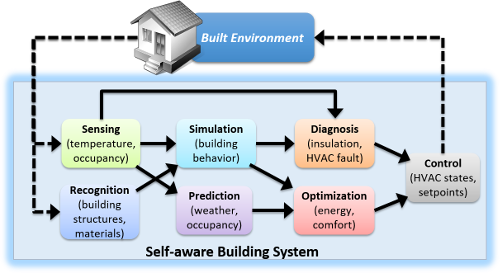}
 \caption{A self-aware building system consists of several autonomous processes.}\label{fig:process}
\end{figure}

On the contrary, {\em self-aware} building systems aim to acquire dynamic faithful knowledge of buildings and learn to control themselves adaptively. Specifically, a self-aware building system can perform the following processes in an autonomous manner \cite{mahdavi2001self}:
\begin{enumerate}[leftmargin=18pt]

  \item {\bf Sensing}:  Gathering environmental data, such as occupants' behavioral information through various sensors, as well as temperature, humidity and external weather information. Its purpose is to acquire the knowledge of the current system state of a building.

  \item {\bf Recognition}: Capturing the physical model of a building by measuring the structures and recognizing the materials. Furthermore, the configurations of HVAC system may be inferred. Its purpose is to create an accurate physical building model.

  \item {\bf Simulation}: Simulating the dynamics of system behavior (e.g., thermal flow behavior) under various parameters. The accuracy of simulation can be improved by a better physical building model and extensive knowledge of the system state. The simulation should be performed in real-time.

  \item {\bf Prediction}: Forecasting future environmental data, such as occupants' behavior and external weather. 

  \item {\bf Optimization}: Deciding the system actions to maximize certain desired objectives (e.g., energy efficiency, comfort). 

  \item {\bf Diagnosis}: Identifying abnormal behavior and adaptively correcting system states. 

  \item {\bf Control}: Applying the actuations of systems in real time.  

\end{enumerate}
These processes interact with each other in a way as visualized in Fig.~\ref{fig:process}. A fully self-aware building system is expected to provide autonomy to these processes, with minimal human assistance. Although automating some of the processes (e.g., control and optimization) have been studied extensively, other processes (e.g., recognition) pose new challenges. This paper aims to shed light particularly on automated recognition for self-aware buildings.

\subsection{Recognition for Self-aware Buildings}
To enable self-aware buildings, it is desirable to acquire the following aspects of information of a building:
\begin{enumerate}[leftmargin=18pt]

  \item {\em Structures}:  Including the dimension and geometry of the building interior.

  \item {\em Materials}:  Including the types of walls (e.g., cement, wood) and windows (e.g., shaded or not), as well as insulation and conductance properties of built materials.

  \item {\em HVAC Systems}:  Including the capacity of heating/cooling power and locations of HVAC ducts.

  \item  {\em Geography}:  Including orientation, environmental shading of the building.


\end{enumerate}
  While the aforementioned information can be provided by the explicit assistance of users {\color{black}(e.g., constructing 3D model from the {\em BIM} models or {\em AutoCAD} diagrams of the building, etc.)}, we aim to develop systems that implicitly gather the information with minimal human assistance.

\section{Augmented Reality}

\begin{figure*}[htb]
 \centering
 \includegraphics[width=1.0\linewidth]{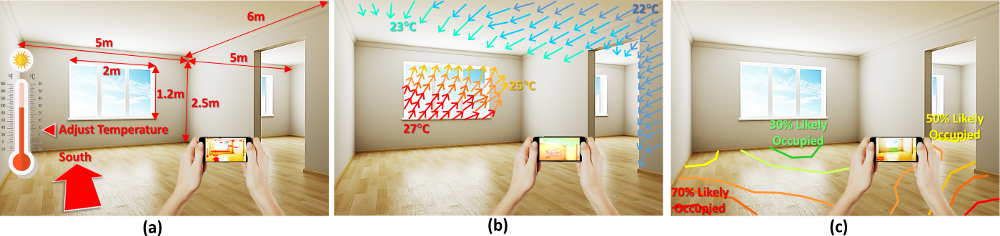}
 \caption{Augmented reality enabled building management system: (a) building geometry recognition, (b) building simulation visualization, and (c) incorporation of user feedback on occupancy behavior.}
 \label{fig:ar_uis}
\end{figure*}

This paper explores the possibilities of using augmented reality, an interactive technology with the physical world by overlaying digital information, as an enabling tool for self-aware buildings. The extensive user-environment interactions in augmented reality allow us to implicitly capture the structures of physical buildings accurately, which can be used for real-time building simulations.

To illustrate the concept, we present several user interfaces of an augmented reality enabled building management system in Fig.~\ref{fig:ar_uis}. For example,  the system will automatically recognize the dimension and geometry of building interior, as visualized in Fig.~\ref{fig:ar_uis}(a). It will identify the locations and dimensions of windows and doors. Also, it will detect the orientation of a building by magnetic sensing.  Hence, a faithful representation of building will be acquired for the construction of physical building model. On the other hand, to increase users' awareness of the energy efficiency of HVAC units, the system will overlay simulated thermal flow information with the physical view of the building, as visualized in Fig.~\ref{fig:ar_uis}(b). Furthermore, users are able to input and modify the information used by building management system through an intuitive interactive user interface. For example, users can provide feedback on occupancy behavior, as visualized in Fig.~\ref{fig:ar_uis}(c).  User input information can improve the system knowledge, and enable a personalized system to address the needs of specific users. 

While there are other options to capture physical building models (e.g., LIDAR), augmented reality is more convenient and cost-effective, providing rich user-building interaction opportunities.

\subsection{Project Tango Platform}

Augmented reality has been a mature technology, with an increasing number of available platforms and commercial products, and emerging new developments in software systems and hardware sensors. Among the state-of-the-art augmented reality platforms, Project Tango is a popular option developed by Google, which will be deployed extensively on enhanced Android tablets or smartphones with depth sensors that are able to acquire 3D information of a physical world in indoor built environments \cite{googletangotablet,lenovotangophone,asustangophone}.

\begin{figure}[htb]
  \centering
  \includegraphics[scale=1.1]{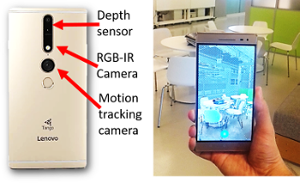}
  \caption{Tango phone (Lenovo Phab pro 2), and capturing 3D scene using Tango phone.}
  \label{fig:tango_phone} 
\end{figure}

Project Tango platform integrates several technologies of augmented reality, such as depth-sensing based on infra-red light, motion tracking camera, and environmental landmark learning \cite{diakite2016first}. We tested on the first commercial Tango phone (Lenovo Phab pro 2 shown in Fig.~\ref{fig:tango_phone}), which shows good accuracy of capturing 3D structures in indoor built environments. A Tango phone provides a collection of pre-installed mobile apps for typical augmented reality experiences, such as Measure app (for remote measurement of object dimensions),  Constructor app (for capturing 3D scenes) and several augmented reality games. Google also provides specified Tango SDK for advanced augmented reality app development.


\begin{figure*}[htbp!]
  \centering
  \includegraphics[width=\textwidth]{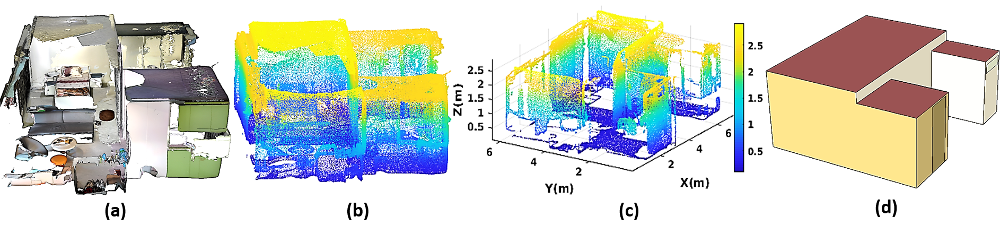}
  \caption{Recognizing building structures from a Tango phone. (a) Captured 3D scene of a room. (b) Converted 3D point cloud. (c) Detected room geometry (e.g., floor and walls). (d) Converted 3D model for building simulation in EnergyPlus.}
  \label{fig:struct_recog}
\end{figure*}


\section{System Prototype}

In this section, we present a system prototype incorporating augmented reality into building management system. In particular, we demonstrate our system for creating an accurate 3D building model by recognizing the building structures, and then inferring the building material properties and certain HVAC system parameters.

\subsection{Recognizing Building Structures}

We first capture the 3D scene of a room by Tango Constructor app, which is able to capture approximate building spatial structures and texture information (depicted in Fig.~\ref{fig:struct_recog}(a)). Then, we convert the 3D scene to a 3D point cloud dataset (depicted in Fig.~\ref{fig:struct_recog}(b)) for advanced processing in MATLAB, which has extensive  Computer Vision System Toolbox. To obtain structural information (i.e., planes corresponding to walls and floor) in the 3D point cloud dataset, we employ the M-estimator Sample Consensus (MSAC) algorithm \cite{msac}. MSAC is a modified version of the Ransom Sample Consensus (RANSAC) \cite{ransac} algorithm for improved surface detection. MSAC is available in MATLAB as function \emph{pcfitplane()}. 

The result of plane detection by MSAC on point cloud dataset is depicted in Fig.~\ref{fig:struct_recog}(c). Initially, MSAC extracts the floor dimensions by detecting the horizontal plane in the point cloud. Then, MSAC is applied iteratively to the point cloud dataset until all vertical planes are detected. Each vertical plane whose height is within a certain threshold of the maximum detected height is labeled as a wall. The x-axis, y-axis, and z-axis respectively represent the length, width, and height of the room that were detected by MSAC.  Finally, the extracted geometry will be used to construct a 3D model for building simulation in EnergyPlus (depicted in Fig.~\ref{fig:struct_recog}(d)). 

\subsection{Inferring Building Materials \& HVAC}


Unlike the physical structures of a building that can be visually perceptible, the building materials are more difficult to be discerned. Although one may use advanced techniques like machine learning and image recognition, we employ a simpler method in this paper. With the accurate 3D models of buildings constructed based on augmented reality technology, it is possible to infer the building material properties and certain HVAC system parameters based on iterative comparisons between simulations (with appropriate calibrations of simulation model parameters) and empirical observations of the real-world building behavior \cite{AFTAB2017141}.

The basic idea is that we first identify a collection of key parameters in the building simulator (see  Table~\ref{tab:recog_params}), and then calibrate these parameters to match as close as possible with the observed thermal response measurements under various external weather conditions and HVAC operations. We employ EnergyPlus \cite{eplus}, a popular building simulation program.
{\color{black} EnergyPlus provides detailed building models, incorporating building structures (e.g., locations of doors and windows, etc.) and material properties, which makes them easier to calibrate than first-principle or non-linear models.}
By iterative calibration of the parameters of building simulation model, we seek to minimize the discrepancy between the actual and simulated indoor temperature values. The calibration of simulation model is performed as follows:
\begin{enumerate}[leftmargin=18pt]
  \item The calibration algorithm sets the initial material properties and HVAC system parameters.
  \item EnergyPlus runs a simulation with adjusted parameters and returns simulated indoor temperature.
  \item The algorithm tracks the changes in the discrepancy between actual and simulated temperature and applies gradient descent to adjust the respective parameters. The process repeats iteratively until the discrepancy is within a certain threshold.
\end{enumerate} 
A final calibrated building model will provide an estimation of the building material properties and certain HVAC system parameters.  This calibration process is carried out using co-simulation, where the EnergyPlus building model is coupled with a calibration algorithm implemented in \emph{Python} using BCVTB (Building Controls Virtual Test Bed) as software interface \cite{bcvtb}.
{\color{black} It is worth noting that gradient descent may arrive at parameter values that agree with the current set of observations but these might not be the true values of the parameters. For this reason, the calibration algorithm continuously checks the model in the background, and re-calibrates it whenever the discrepancy between the actual and simulated temperature exceeds the threshold. This way, the model always provides a good estimation of the parameter values.}
\begin{figure}[htbp!]
\centering
\includegraphics[width=\linewidth]{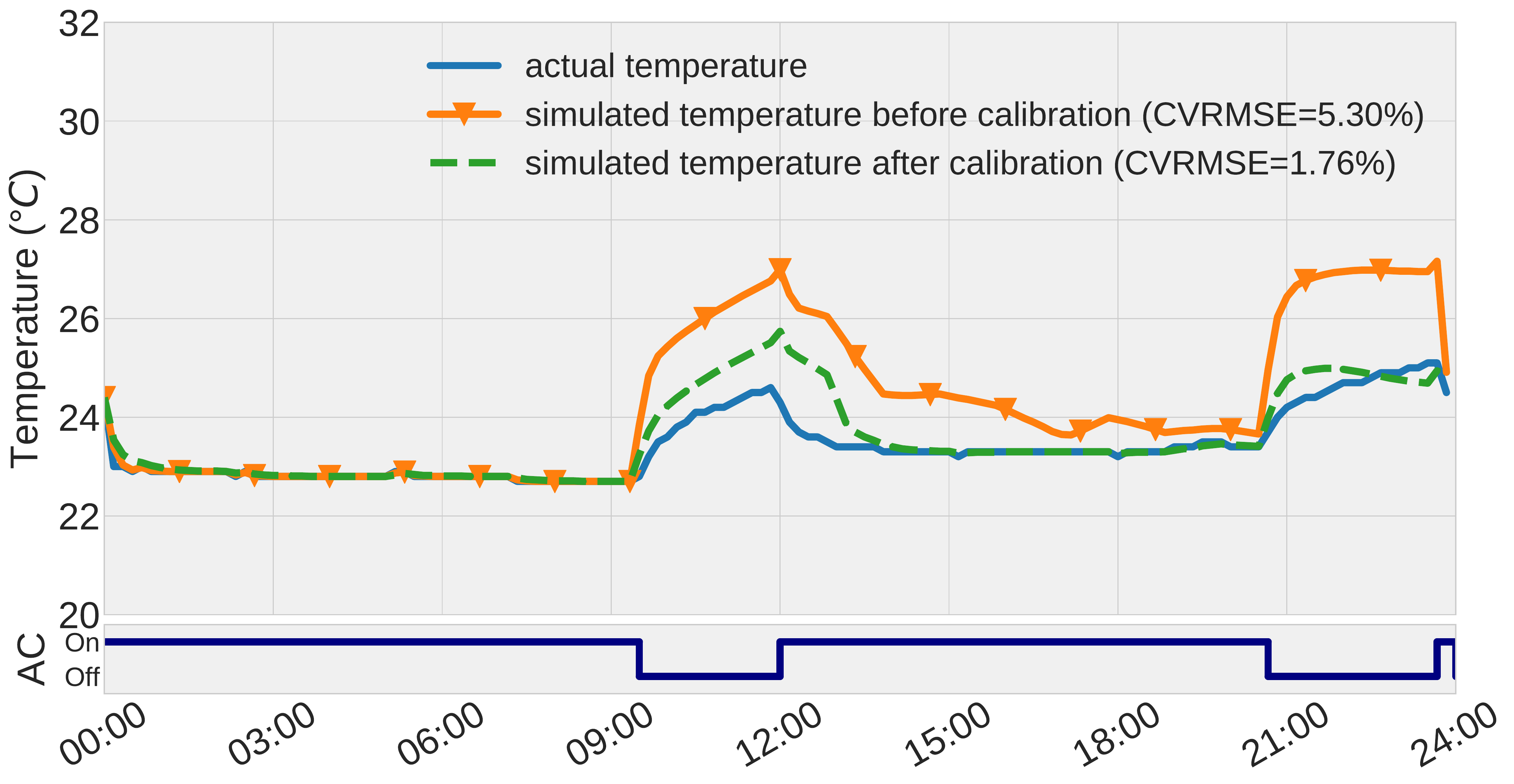}
  \caption{Inferring building material properties and HVAC system parameters based on calibration of building simulations.}
\label{fig:sim_model_calib}
\end{figure}
A demonstration using the 3D model from Fig.~\ref{fig:struct_recog}(d) is illustrated in Fig.~\ref{fig:sim_model_calib}, showing the thermal response of the initial uncalibrated model and of the iteratively calibrated model. After calibration, the simulated temperature of the calibrated EnergyPlus model can closely match the actual temperature. Fig.~\ref{fig:sim_model_calib} also depicts the observed status of HVAC system during different times of the day, as in the bottom strips. Finally, Table~\ref{tab:recog_params} presents the inferred parameter values of building simulation model, including building material properties and HVAC system.

\begin{table}[htb!]
  \centering
  \scalebox{0.85}{
  \setlength\tabcolsep{12pt}
  \begin{tabular}{c |  c |  c } \hline \hline
    Parameter & Field & Calibrated Value \\  \hline \hline

    \multirow{2}{*}{Walls} & Thickness    & 30.0cm \\
                           & Conductivity & 0.311W/(m-K) \\\hline
    \multirow{2}{*}{Windows} & Thickness  & 0.31cm \\
                                          & Conductivity & 0.85W/(m-K) \\\hline
    \multirow{2}{*}{Door} & Thickness & 2.54cm \\
                                      & Conductivity & 0.15W/(m-K) \\\hline
    \multirow{2}{*}{Roof} & Thickness & 10.16cm \\
                          & Conductivity & 0.53W/(m-K) \\\hline
    \multirow{2}{*}{HVAC} & Cooling Capacity	& 8943W \\ 
                          & Air Flow Rate & 0.384m$^3$/sec \\ \hline\hline
   \end{tabular}}
   \caption{Key building simulation model parameters.}
   \label{tab:recog_params}
\end{table}

\section{Conclusion}
This paper first presented a notion of self-aware building that is capable of recognizing of building  structures and properties to construct accurate simulation models. Then, we demonstrated a system prototype using augmented reality to capture the build structures and then to infer building materials and HVAC system parameters. In future work, we will implement a full system that incorporates simulation visualization, user feedback elicitation, and automatic real-time simulation-guided control for HVAC operations.

\bibliographystyle{IEEEtran}
\bibliography{refs}

\end{document}